\documentstyle[12pt,epsf,epsfig]{article}
\def\be{\begin{equation}}
\def\ee{\end{equation}}
\def\ba{\begin{eqnarray}}
\def\ea{\end{eqnarray}}
\def\ga{\mathrel{\raise.3ex\hbox{$$\kern-.75em\lower1ex\hbox{$\sim$}}}}
\def\la{\mathrel{\raise.3ex\hbox{$<$\kern-.75em\lower1ex\hbox{$\sim$}}}}

\begin{document}

\begin{titlepage}
\pagestyle{empty}
\baselineskip=21pt
\rightline{TPI--MINN--01/38}

\rightline{August 2001}
\vskip.25in
\begin{center}

{\large{\bf Phenomemology of a Realistic Accelerating Universe\\[1mm]
Using Tracker Fields}}
\end{center}
\begin{center}
\vskip 0.4in

{Vinod B. Johri*}
\vskip 0.2in
{\it
{Theoretical Physics Institute, School of Physics and Astronomy, \\
University of Minnesota, Minneapolis, MN 55455, USA}}
\vskip 0.5in
{\bf Abstract}
\end{center}
\baselineskip=18pt \noindent
%%%%%%%%%%%%%%%%%%%%%%%%%%%%%%%%%%%%%%%%%%%%%%%%%%%%%%%%%%%%%%%%%%%%%
We present a realistic scenario of tracking of scalar fields with varying equation of state. The astrophysical constraints on the evolution of scalar fields in the physical universe are discussed. The nucleosynthesis and the galaxy formation constraints have been used to put limits on $\Omega_\phi$ and estimate $\epsilon$ during cosmic evolution. Interpolation techniques have been applied to estimate $\epsilon\simeq0.772$ at the present epoch. The epoch of transition
from matter to quintessence dominated era and consequent onset of acceleration in cosmic expansion is calculated and taking the lower limit  $\Omega_n^0 = 0.2$ as estimated from 
$SN_e I_a$ data
, it is shown that the supernova observations beyond redshift $z=1$ would reveal
deceleration in cosmic expansion
\\[3mm]
%%%%%%%%%%%%%%%%%%%%%%%%%%%%%%%%%%%%%%%%%%%%%%%%%%%%%%%%%%%%%%%%%%%%
\vspace*{18mm}
\begin{flushleft}
\begin{tabular}{l} \\ \hline
{\small Emails: vinodjohri@hotmail.com}
\end{tabular}
\end{flushleft}

\end{titlepage}
%\newpage
\baselineskip=18pt
%%%%%%%%%%%%%%%%%%%%%%%%%%%%%%%%%%%%%%%%%%%%%%%%%%%%%%%%%%%%%%%%%%%%%

\section{Introduction}
The observational view of the universe has drastically changed during the last ten years. Until a decade ago, the universe was supposed to be matter dominated and the cosmic expansion was understood to be slowing down; consequently the Einstein de Sitter model was taken to be the standard model of the observable universe. However, the recent studies undertaken by the Supernova Cosmology Project Team \cite{1}  and the High Redshift Search Team \cite {2} reveal that the distant supernovae are fainter and thus more distant than expected for a decelerating universe. It implies that the rate of cosmic expansion is accelerating. This  provides empirical evidence of the existence of an exotic matter with repulsive character in the universe which counters  gravitational attraction of ordinary matter. If the energy density of such exotic matter is dominant over that of ordinary matter, it may cause acceleration in cosmic expansion.
The hypothesis of the existence of exotic matter is in fact supported by the analysis of the luminosity magnitude and redshift measurements of type$I_a$ Supernovae   that leads to the estimate
                  $\Omega _n = 0.3\pm0.1$ for ordinary matter and
	$\Omega_X = 0.7\mp0.1$ for exotic matter
There are various plausible candidates for the exotic matter, sometimes referred to as dark energy. The Einstein's Cosmological constant $\Lambda$ is a natural choice owing to its repulsive character but it is not acceptable due to its extremely low magnitude compared to particle physics scales. A dynamical cosmological constant with negative pressure looks promising and recently a lot of research has been undertaken to find scalar fields with suitable potentials which roll down slowly to give rise to negative pressure during later stages of evolution and behave like $\Lambda_{effective}$. Any physical phenomenon, which may cause negative pressure and violate the strong energy condition, would support cosmic acceleration and may be called Quintessence.  Various scalar potentials with evolving equation of state have been proposed as  probable candidates  for quintessence. These include inverse power law, exponential and hyperbolic potential functions \cite{3}-\cite{12} of the scalar field. The growth of scalar fields in the universe is severely restricted by the observational constraints; for instance, the magnitude of $\Omega_\phi $ must be low enough so as not to disturb the observed helium abundance at nucleosynthesis epoch$(z=10^{10})$ and not to  interfere with the process of formation of galactic structure $(2<z<4)$ but at the same time, the scalar field must be dominant over matter/radiation at the present epoch in order to satisfy the observational estimate of $\rho_\phi\sim 2\rho_m.$ . This demands extreme fine-tuning of scalar fields relative to matter field. To get rid of fine tuning, the  idea of 'tracking ' was put forward \cite{13,14} and attractor-type potentials were investigated which, irrespective of their divergent initial conditions, evolve along a common track   with the background energy density of matter and radiation to end up in the domination of $\rho_\phi$ over $\rho_n$. Whereas, tracking solves the fine tuning and coincidence problems, it does not ensure the compatibility of the scalar field with the observational constraints. Since there is no control over the slow roll down and the rate of growth of the scalar field energy during tracking, the transition to the scalar field dominated phase may take place much later than observed.

The relative growth of the scalar field energy, represented by $\Omega_\phi$, may be controlled by monitoring the evolution of the equation of state of the scalar fields through the tracking parameter $\epsilon$   at various observational landmarks in the thermal history of the universe. With this perspective, the concept of 'integrated tracking' was introduced \cite {15} , which links tracking to  the estimated value of  $\epsilon$ at various epochs during cosmic evolution. The choice of the desired tracker potential is thus narrowed down to those which satisfy the established tracking criterion \cite {15} with $\epsilon$ conforming to the estimated values derived from the observational constraints. This fixes the path of evolution of the tracker field subject to  the errors in estimation of $\epsilon$. As shown in this paper $\Omega_\phi$ is a function of $\epsilon$ and redshift $z$. Knowing that at the point of transition from matter dominated to the scalar field dominated era $\Omega_\phi = 0.5, \epsilon=0.666$ and  $z=0.414$, we can interpolate the value of  $\epsilon$ at other points in cosmic evolution and plot the path of evolution of the tracker field, independent of the choice of tracker potential. In fact, it may not be possible to find a unique tracker potential to cover the entire tracking range but scalar potentials to display tracker behavior under restrictive assumptions may be given.In our previous paper \cite {15} , we have traced the  paths of tracker potential under the assumption that  $\epsilon=constant$  throughout tracking. But it is unrealistic since the value of  $\epsilon$ goes on diminishing as you go back in time to the early universe.
We present a realistic account of the behavior of tracker potential in this paper by starting with the precisely known values of $\Omega_\phi$ and $\epsilon$ at $z= 0.414$ and use interpolation methods to estimate  $\epsilon=0.772,\Omega_\phi  = 0.66 $ at the  present epoch ($z=0$) . The astrophysical constraints on  $\Omega_\phi$ briefly outlined in \cite {15} , have been re-examined in this paper and variation of the tracking parameter, the cosmological density 
parameter and the equation of state of the scalar field  has been plotted in terms of redshift. 

\section{ Dynamics of Tracking}

Let us first consider CDM cosmology with Quintessence -- the rolling
scalar fields, with evolving equation of state, which acquire
repulsive character (owing to negative pressure) during the late evolution
of the universe. The quintessence in the present day
observable universe, behaves like $\Lambda_{eff}$ and may turn out to be
the most
likely form of dark energy which induces acceleration in the cosmic
expansion.

Consider the homogeneous scalar field $\phi(t)$ which interacts with matter
only through gravity. The energy density $\rho_\phi$ and the pressure
$p_\phi$
of the field are given by
\be
\rho_\phi\, = \frac 12\,\dot\phi^2 + V(\phi)
\ee
\be
p_\phi\, = \frac 12\,\dot\phi^2 - V(\phi)
\ee
The equation of motion of the scalar field
\be
\ddot{\phi} + 3H\dot{\phi} + V'(\phi)\, = 0, \qquad
V'(\phi)\equiv\frac{dV}{d\phi}
\ee
leads to the energy conservation equation
\be
\dot{\rho}_\phi + 3H(1+w_\phi)\rho_\phi\, = 0
\ee
where $w_\phi\equiv\frac{p_\phi}{\rho_\phi}$ and
$H\equiv\frac{\dot{a}}{a}$
is
the
Hubble constant. Accordingly, $\rho_\phi$ scales down as
\be
\rho_\phi\,\sim {a}^{-3(1+w_\phi)} , \quad -1\leq w_\phi \,\leq 1
\ee
Obviously, the scaling of $\rho_\phi$ gets slower as the potential energy
$V(\phi)$ starts dominating over the kinetic energy $\frac 12 \dot{\phi}^2$
of the scalar field and $w_{\phi}$ turns negative.

Since there is minimal interaction of the scalar field with matter and
radiation,
It follows from Eq.(4) that the energy of matter and radiation is
conserved
separately as
\be
\dot\rho_n + 3H(1 + w_n)\rho_n = 0
\ee
Accordingly
\be
\rho_n\,\sim a^{-3(1+w_n)}
\ee
where $\rho_n$ is the energy density of the dominant constituent (matter or
radiation) in the universe with the equation of state $p_n = w_n\,\rho_n$
where $w_n = \frac 13$ for radiation and $w_n = 0$ for matter.

Although, the scalar field is non-interactive with matter, it affects the
dynamics of cosmic expansion through the Einstein field equations. Assuming
large scale spatial homogeneity and isotropy of the universe, the field
equations for a flat Friedmann model are
\be
 H^2 \, = \frac{\rho_n + \rho_\phi}{3M_p^2}
\ee
and
\be
 \frac{2\ddot {a}}{a}\, = -\frac{\rho_n+\rho_\phi +3p_n +3p_\phi}{3M_p^2}
\ee
where $M_p =2.4\times 10^{18}$ GeV is the reduced Planck mass.

Denoting the fractional density of the scalar field by $\Omega_\phi\equiv
\frac{\rho_\phi}{\rho_n+\rho_\phi}$ and that of the matter/radiation
field
by
$\Omega_n\equiv\frac{\rho_n}{\rho_n+\rho_\phi}$, equations (8) and (9)
may
be
rewritten as
\be
 \Omega_n + \Omega_\phi\, = \, 1
\ee
and
\be
2\frac{\ddot a}{a} = -\frac{\rho_n}{3M_p^2}\,[(1+3w_n)+(1+3w_\phi)
\frac{\Omega_\phi}{\Omega_n}]
\ee

The relative growth of $\Omega_\phi$ versus $\Omega_n$ during the cosmic
evolution is given by
\be
\frac{\Omega_\phi}{\Omega_n} = \frac{\Omega_\phi^0}{\Omega_n^0}\,
\biggl(\frac{a}{a_0}\biggr)^{3\epsilon}
\ee
where the tracking parameter $\epsilon\equiv w_n - w_\phi$ and
$\Omega_\phi^0$, $\Omega_n^0$ denote the values of $\Omega_\phi$ and
$\Omega_n$
at the present epoch $ (a = a_0)$. Assuming that that $\Omega_\phi^0
\simeq 2\Omega_n^0$, Eq.(12) may be expressed in terms of the
red-shift $z$ as below
\be
2(\Omega_\phi^{-1} - 1) =\, (1+z)^{3\epsilon}
\ee

If we insist that the scalar field, regardless of its initial value, should
behave like $\Lambda_{eff}$ today, it must obey tracking conditions
\cite{13,15} which have
 wide ramifications for quintessence fields already discussed in detail
\cite{13,14,15}. In nutshell, tracking
consists in synchronised scaling of $\rho_\phi$ and $\rho_n$ along a common
evolutionary track so as to ensure the restricted growth of $\Omega_\phi$
during the cosmic evolution in accordance with the observational constraints. 
As discussed in \cite{15}, the tracking
parameter $\epsilon$ plays a vital role in monitoring the desired growth of
$\Omega_\phi$. The existence criterion for tracker fields

 requires
$\epsilon$ to satisfy the condition
$\frac{\epsilon\Omega_n}{2(1+w_\phi)} <1$
and also to conform to cosmological constraints mentioned therein. Here we
investigate the consequent limits on the tracking parameter $\epsilon$
imposed
by these constraints to ensure tracking by a quintessence field.

According to the Eq. (13), $\Omega_\phi$ depends both upon the redshift $z$
and the tracking parameter $\epsilon$. From a general functional form
$\Omega_\phi = f(z, \epsilon)$, it is difficult to map out
$\Omega_\phi -
z$ variation through tracking unless we are able to fix the value of
$\epsilon$ corresponding to known $\Omega_\phi$ at certain points in the
phase
space of $z - \epsilon$. This is achieved with the help of the astrophysical
constraints discussed in this paper. Having evaluated $\Omega_\phi$ at some
typical points $(z_0, \epsilon_0)$, we can interpolate
$\Delta\Omega_\phi$
at
neighbouring points $(z_0 + \Delta z_0 , \epsilon_0 +
\Delta\epsilon_0)$.
Using
this technique, realistic tracking diagrams of $\Omega_\phi - z$ variation,
$w_\phi - z$ variation and
$\epsilon - z$ variation may be drawn as shown in the figures 1,2 and 3.

%%%%%%%%%%%%%%%%%%%%%%
\begin{figure}
\begin{center}
\mbox{\epsfig{file=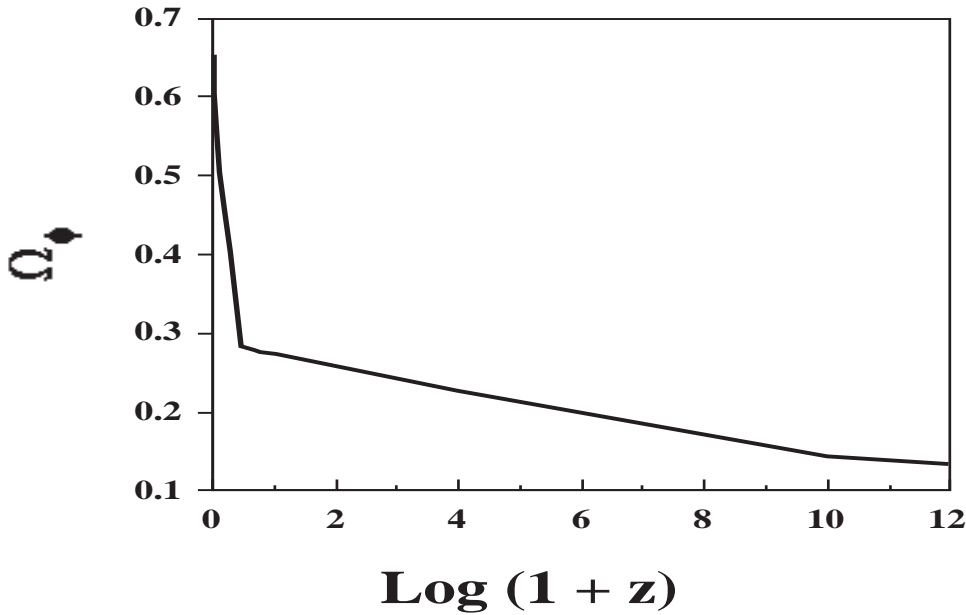,height=8.3cm,width=14cm}}
\end{center}
\caption[.]{\label{fig:bsg1}\it
Variation of $\Omega_\phi$ versus Redshift $z$ in QCDM Cosmology
assuming $H_o=65$ Km/Mpc/s.}
\end{figure}
%%%%%%%%%%%%%%%%%%%%

%%%%%%%%%%%%%%%%%%%%%%
\begin{figure}
\begin{center}
\mbox{\epsfig{file=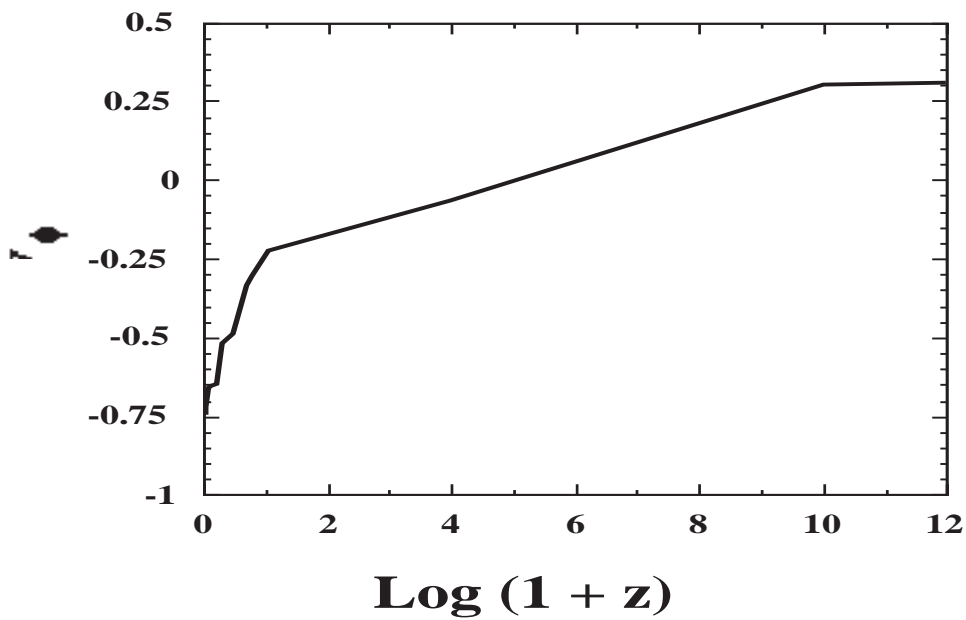,height=8.3cm,width=14cm}}
\end{center}
\caption[.]{\label{fig:bsg2}\it
Evolution of the equation of state of the quintessence field in QCDM
Cosmology
($H_o=65$ Km/Mpc/s).}
\end{figure}
%%%%%%%%%%%%%%%%%%%%

%%%%%%%%%%%%%%%%%%%%%%
\begin{figure}
\begin{center}
\mbox{\epsfig{file=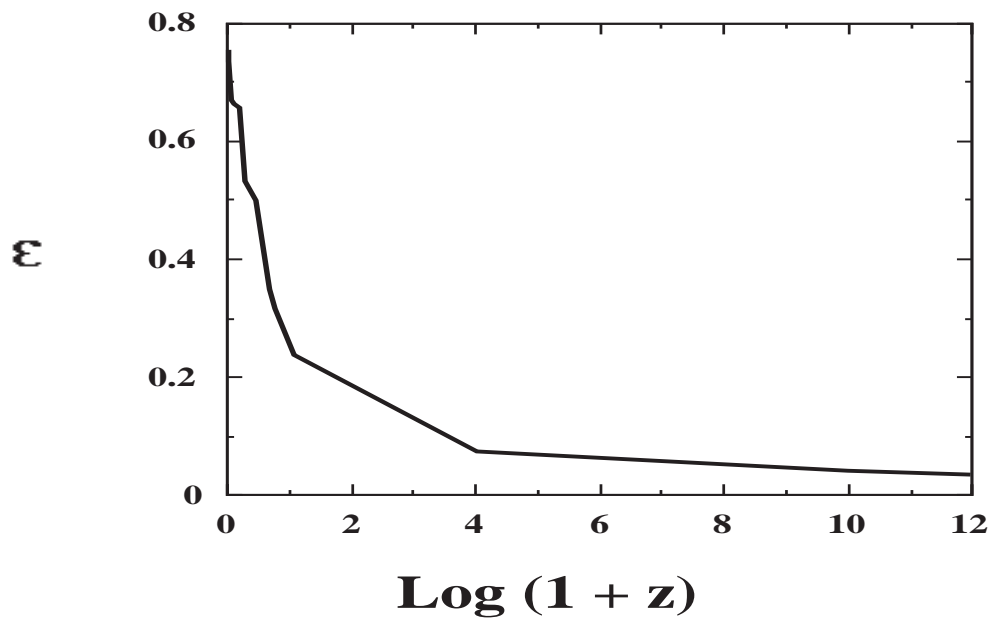,height=8.3cm,width=14cm}}
\end{center}
\caption[.]{\label{fig:bsg3}\it
Variation of the tracking parameter $\epsilon$ versus Redshift $z$
in QCDM Cosmology ($H_o=65$ Km/Mpc/s).}
\end{figure}
%%%%%%%%%%%%%%%%%%%%

In this connection, the following differential relation, derived from
Eq.(13)
\be
 - \frac{\Delta\Omega_\phi}{3\epsilon\Omega_n\Omega_\phi} \, =
\frac{\Delta z}{1+z}
 + \frac{\Delta\epsilon}{\epsilon} \ln(1+z).
\ee
is found quite useful in interpolating the increment
$\Delta\Omega_\phi$ in
terms of the increments $\Delta z$ and $\Delta\epsilon$. It is noteworthy
that the contribution of the term $\frac{\Delta z}{1+z}$ is very small
compared to the contribution of $\frac{\Delta\epsilon}{\epsilon}$ at high
 redshifts.

 Let us now reconsider the astrophysical constraints discussed in our
previous
paper \cite{15}, try to refine them and examine their implications for
quintessence
fields.

{\bf 1. The Nucleosynthesis Constraint.}
 The first constraint on $\Omega_\phi$ during the cosmic evolution
comes
from
the
 helium abundance at the nucleosynthesis epoch $( z\sim 10^{10})$. The
presence
 of an additional component of energy in the form of quintessence
field
with
 energy density $\rho_\phi$ results in an increase in the value of the
Hubble
 constant $H$ as given by the differential of the Friedmann equation
 \be
 \frac{2\delta H}{H} \, = \frac{\delta\rho}{\rho}\, =
\frac{\rho_\phi}{\rho}.
 \ee
 This, in turn, yields a higher ratio of neutrons to protons at the
freeze-out
 temperature (1 MeV) of the weak interactions and a consequent higher
percentage
 of the helium abundance in the universe. Assuming the existence of
three
known
 species of neutrinos, the nucleosynthesis calculation \cite{16} yields
 \be
 \frac{\delta\rho}{\rho} \, = \, \frac{7(N_\nu - 3)}{43}
 \ee

%%%%%%%%%%%%%%%%%%%%%%
\begin{figure}
\begin{center}
\mbox{\epsfig{file=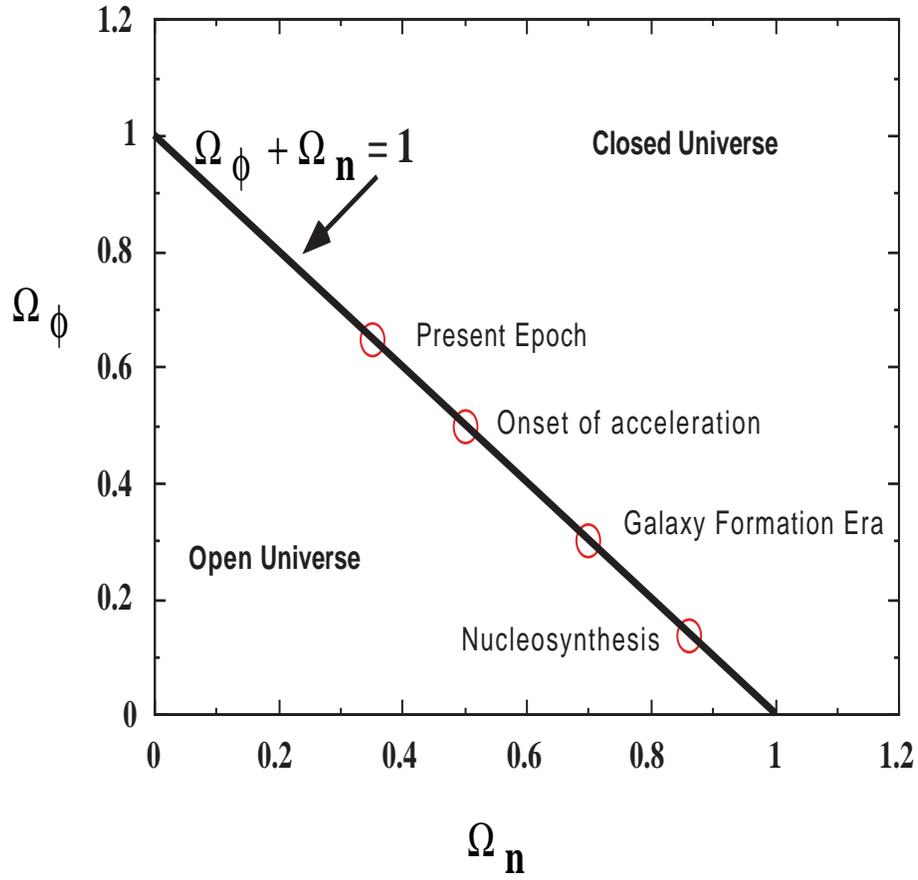,height=12cm,width=12cm}}
\end{center}
\caption[.]{\label{fig:bsg4}\it
The important cosmic events corresponding to the astrophysical constraints
are marked by circles on the thick line (representing spatially flat
universe)
in QCDM Cosmology.}
\end{figure}
%%%%%%%%%%%%%%%%%%%%

 Since the number of neutrino species $N_\nu < 4$, we arrive at the
constraint
 $\frac{\delta\rho}{\rho} < \frac{7}{43}$. If the contribution
$\delta\rho$
 comes from the quintessence field instead, the above constraint
translates
into
 nucleosynthesis constraint on $\Omega_\phi$ as follows
 \be
 \Omega_\phi = \frac{\rho_\phi}{\rho_n + \rho_\phi} < \frac{7}{50} = 0.14
 \ee
 The corresponding value of the tracking parameter is $\epsilon\leq0.035$.

 %%%%%%%%%%%%%%%%%%%%%%%%%%%%%%%%%%%%%%%%%%%%%%%%%%%%%
{\bf 2. Galaxy Formation Constraint.}
 According to the current estimates in CDM cosmology, the galactic
structure
 formation takes place between the redshift $z = 4$ and $z = 2$. The
clumping
 of matter into galaxies demands the dominance of gravitational
attraction
 during this period. Therefore, the repulsive force of quintessence
must
be
 relatively weak and $\Omega_\phi$ must be reasonably less than 0.5
during
 the galaxy formation era. Interpolation from Eqs. (13) and (14)
shows
that
 $0.33\leq \epsilon < 0.5$ during the galaxy formation era.

{\bf 3. Present Epoch.}
 Two recent surveys \cite{1,2} based on $SN_e I_a$ measurements
 predict accelerating cosmic expansion with $\Omega_\phi\simeq 0.7\
 \pm 0.1$ at the present epoch (z=0). For the sake of computation , 
 we take the value $\Omega_\phi^0 = 0.66$ at $z=0$ and interpolate
 the values at other redshifts and plot the variation of $\Omega_\phi$,
 $w_\phi$ and $\epsilon$ in terms of $z$ as shown in figures 1,2 and 3.
  
 The constraint $\ddot a> 0 $
inserted
 in Eq. (11) leads to
 $\epsilon > 0.5$ at the present epoch. Interpolation with
 the help of Eq.(14) places $\epsilon$ around 0.77.

{\bf 4. Transition to Accelerated Expansion (Quintessence Dominated Era).}
 The onset of acceleration( $\ddot a \geq 0$) in the observable universe
takes place around
 the value of $\Omega_\phi\geq 0.5$ which corresponds to
 $\epsilon\sim0.666$
 from Eq.(11), at a redshift of $z= 0.414$ (assuming $\Omega_n^0 =0.33$)from Eq.(13).

In figure 4, the cosmic events are depicted sequentially by the thick
line in
the $\Omega_n - \Omega_\phi$ diagram. The events corresponding to the above
constraints
are marked by circles.
%%%%%%%%%%%%%%%%%%%%%%%%%%%%%%%%%%%%%%%%%%%%%%%%%%%%%%%%%%%%%%%%%%%
\section{Transition to scalar Field Dominated Era}

As revealed by supernova observations, we are living in the quintessence dominated universe 
today with $\Omega_\phi\simeq  2\Omega_n$. When did the transition from the matter to 
quintessence(scalar field ) dominated era occur? According to tracker field theory, the 
transition occurs at a stage when $\rho_\phi\geq \rho_n$ or $\Omega_\phi\geq 0.5$ during
cosmic evolution. Since $\epsilon = 0.666$ at this stage, the transition takes place at
a redshift $z^*$ as derived from Eq.(12)

\be
        z^*  \, = \, \sqrt{\frac{\Omega_\phi^0}{\Omega_n^0}}\, -\,1
\ee

Assuming that the onset of acceleration  starts at the transition epoch, the galaxies 
beyond redshift $z^*$ will be found to be decelerating. 
 Taking the lowest value $\Omega_n^0 = 0.2$ out of the estimated range from the $SN_e$
 observations, the largest redshift for transition from Eq.(18) comes out to be $z^*=1$.
 The recent discovery \cite{17}of a type $I_a$ supernova at a redshift $z\sim 1.7$ from
 Hubble Deep Field observations shows the first glimpse of cosmic deceleration It is 
 consistent with the
 tracker field theory which predicts that the galaxies beyond $z=1$ would have 
 decelerating expansion. This  enables us to make another prediction that the future 
 supernovae observations
 should be able to determine the point of onset of cosmic acceleration and with the 
 knowledge of this redshift, it should be possible to fix the value of $\Omega_n^0$ and map 
 out the exact course of the future evolution of the universe.

 To conclude this paper, we would like to point out that the rolling scalar fields, minimally
 coupled to matter, seem to be most favourable candidate for the exotic matter (dark energy)
 in the universe and their behaviour can be completely investigated in the framework of 
 tracker field theory. We have discussed, in this paper, the variation of $\Omega_\phi$, the 
 tracking parameter $\epsilon$ and the equation of state parameter $w_\phi$ as function of the
  redshift $z$. Likewise, we can investigate the relative scaling of the energy densities     $\rho_\phi$ and $\rho_n$ as function of $z$ during cosmic evolution, using Eqs. (7) and (13).

{\large\bf Acknowledgments}
This work was supported  by UGC grant from India. The author
acknowledges useful
discussions and valuable help of Keith Olive and Panagiota Kanti and
hospitality of
Theoretical Physics Institute, University of Minnesota.

* Permanent Address: Department of Mathematics and Astronomy, Lucknow
University,
 Lucknow 226007. India.

\end{document}